\documentclass[9pt,twocolumn,twoside]{opticajnl}
\journal{opticajournal} 

\setboolean{shortarticle}{false}

\title{Enhancing analytical photon transport in turbid media through physics-guided neural network corrections}

\author[1]{Guojun Tang}
\author[1]{Paul D. Teal}
\author[1,*]{Alexander Doronin}

\affil[1]{Victoria University of Wellington, School of Engineering and Computer Science, Wellington, New Zealand}

\affil[*]{alex.doronin@vuw.ac.nz}

\begin{abstract}
Analytical methods provide rapid and computationally efficient predictions of light transport in turbid media, making them attractive for real-time and inverse applications. However, their accuracy is limited by the approximations and assumptions used to obtain tractable solutions. On the other hand, Monte Carlo (MC) simulation is widely regarded as the "gold standard" for modeling light transport in turbid media, but its high computational cost limits its use in applications requiring repeated forward evaluations. Here, we introduce a physics-guided residual learning framework that combines the computational efficiency of analytical modeling with the accuracy of MC simulation. As a representative case study, we consider the diffusion approximation (DA), which can exhibit systematic errors under relatively strong absorption, weak scattering, and imperfect boundary treatment. Given the absorption coefficient, scattering coefficient, and anisotropy factor, the corresponding DA prediction is incorporated as a physical prior, while the network learns only the residual between the analytical prediction and the MC reference. Across the evaluated optical-property range, the proposed framework reduces scalar-reflectance RMSE and spatially resolved reflectance RMSLE by factors of approximately 62 and 14, respectively, relative to the diffusion approximation. With only 1\% of the available training data, it achieves approximately 87\% and 93\% lower corresponding errors than a direct neural-network surrogate. These results demonstrate that learning corrections to an analytical physical baseline improves both prediction accuracy and data efficiency.
\end{abstract}

\setboolean{displaycopyright}{false} 

\begin{document}

\maketitle

\section{Introduction}
Accurate modeling of light transport in biological tissues is a fundamental problem in biomedical imaging, computer graphics and light-based sensing as it determines how optical radiation interacts with tissue and enables physiological and morphological information to be inferred from measured signals~\cite{tuchin2016handbook}. In turbid media, such as biological tissues, photons undergo repeated scattering and absorption events governed by the tissue optical properties, including the absorption coefficient $\mu_\text{a}$, scattering coefficient $\mu_\text{s}$, scattering anisotropy $g$, refractive index $n$, and geometrical characteristics~\cite{wang2007biomedical,hielscher2016forward}. The radiative transfer equation (RTE) provides the fundamental theoretical framework for describing this process. However, general analytical solutions are difficult to obtain because the RTE couples spatial and angular variables through an integral scattering term. Consequently, photon-transport modeling has traditionally developed along two broad computational directions: deterministic methods, which seek tractable analytical or numerical solutions or approximations to the RTE, and stochastic methods, which simulate photon transport statistically~\cite{Bykov2017LightPI}.

Among deterministic approaches, the diffusion approximation (DA) remains one of the most widely used because of its analytical tractability and low computational cost. By assuming that the radiance field becomes nearly isotropic after sufficient multiple scattering, DA provides closed-form or semi-analytical descriptions that are particularly attractive for diffuse reflectance modeling and inverse estimation of tissue optical properties~\cite{wang2007biomedical,farrell1992diffusion,Martelli2010Light}. This computational efficiency, however, comes with a restricted validity regime. Systematic deviations arise when absorption is appreciable relative to reduced scattering, propagation distances are insufficient for the radiance to become diffuse, or boundary conditions are inadequately represented. For pencil-beam illumination, additional errors arise from replacing anisotropic transport with an equivalent isotropic description and approximating the incident beam by an embedded isotropic source~\cite{Wang2000Sources}. Boundary treatment is particularly important: Haskell et al. demonstrated that neglecting the surface boundary can substantially bias the optical coefficients recovered from diffuse optical measurements~\cite{haskell1994boundary}. 

\begin{figure*}[t]
\centering
\begin{minipage}[t]{0.45\textwidth}
    \centering
    \includegraphics[width=0.8\textwidth]{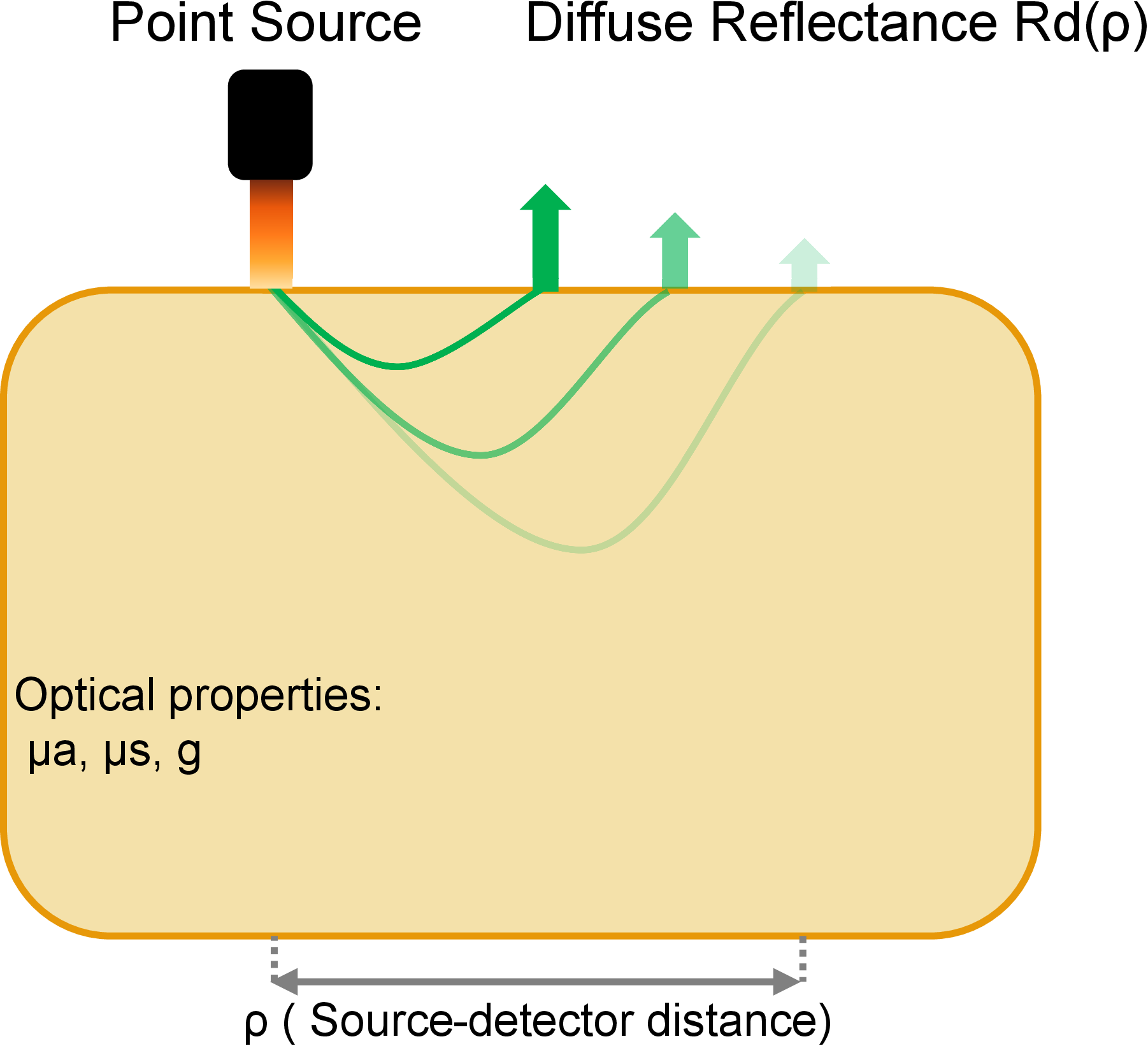}
    \vspace{1mm}
    \makebox[\textwidth][c]{\small (a)}
\end{minipage}\hfill
\begin{minipage}[t]{0.45\textwidth}
    \centering
    \includegraphics[width=0.9\textwidth]{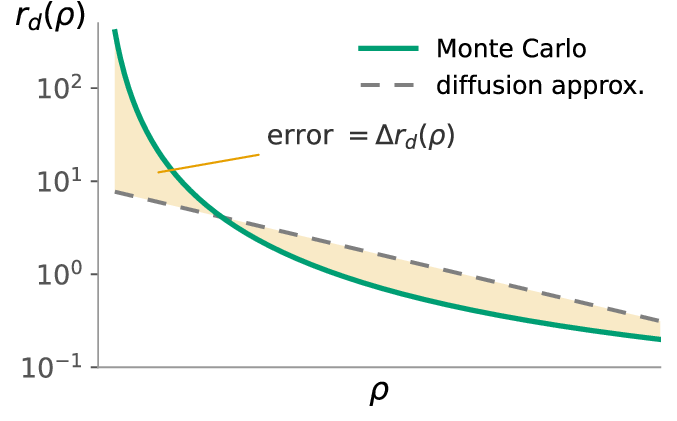}
    \vspace{1mm}
    \makebox[\textwidth][c]{\small (b)}
\end{minipage}
\caption{(a) Spatially resolved diffuse reflectance $R_\text{d}(\rho)$ in a medium with optical properties $(\mu_\text{a}, \mu_\text{s}, g)$. 
(b) Conceptual illustration of the discrepancy between the diffusion approximation (dashed) and Monte Carlo (solid). The shaded gap, $\Delta R_\text{d}(\rho)$, represents the correction learned by the proposed framework.}
\label{fig:concept}
\end{figure*}

These limitations are also supported by the experimental work of Yoo, Liu, and Alfano, who observed departures from diffusion theory at propagation distances of only several transport mean free paths, with stronger deviations as scattering anisotropy increased~\cite{Yoo1990WhenDT}. Such inaccuracies are especially consequential in inverse problems, where systematic errors in the forward model can propagate into biased estimates of tissue optical properties~\cite{arridge2009optical}. Together, these limitations motivate approaches that retain the analytical structure of DA while correcting its regime-dependent discrepancies from more accurate transport models.

At the other end of the modeling spectrum, stochastic approaches such as Monte Carlo (MC) simulation are widely regarded as the gold standard for photon transport in turbid media. MC directly simulates photon propagation through random sampling of scattering, absorption, and boundary interactions, and a comprehensive overview of the historical development of MC modeling in biomedical optics is presented in~\citep{jacques2022history}. Unlike DA, MC does not rely on the near-isotropic-radiance assumption and can therefore be applied over a much broader range of optical regimes. Its principal limitation is computational cost: accurate estimates of diffuse reflectance may require very large numbers of photon packets to suppress stochastic noise. This creates a long-standing practical gap between fast but approximate analytical methods and accurate but computationally demanding stochastic simulations.

More recently, machine learning has emerged as a third computational strategy for photon transport, offering the possibility of learning fast surrogate mappings from optical properties to measurable quantities such as diffuse reflectance~\cite{nguyen2025neuralrte}. Conventional neural networks can be trained to predict reflectance directly from inputs such as $\mu_\text{a}$, $\mu_\text{s}$, and $g$, thereby approximating the behavior of high-fidelity MC simulations at substantially reduced inference cost. However, purely data-driven surrogates typically learn the full input-output relationship from data alone, may require large training datasets, and do not explicitly exploit the analytical structure already available from established transport models. As a result, a considerable amount of known physical information may remain unused during learning~\cite{Wang2021PhysicsGuidedDL,daw2022physics}.

On the other hand, physics-informed neural networks have found widespread use in practical applications across a broad range of spatial and temporal scales, including both forward and inverse problems~\cite{Wang2021PhysicsGuidedDL}. By incorporating governing equations and physical constraints into the learning process, these networks provide a flexible framework for integrating data with established physical models. 

In this work, we address the long-standing challenge of accurately and efficiently solving the \textbf{forward photon transport problem} by introducing a physics-guided residual learning (PGRL) framework that combines the computational efficiency of the diffusion approximation with the accuracy of Monte Carlo (MC) supervision. We consider the prediction of both scalar diffuse reflectance $R_\text{d}$ and spatially resolved diffuse reflectance $R_\text{d}(\rho)$. As illustrated in Figure~\ref{fig:concept} (a), incident light propagates through a medium characterized by the optical properties $(\mu_\text{a},\mu_\text{s},g)$, and following multiple scattering and absorption, a fraction of the radiation re-emerges from the surface to form the spatially resolved reflectance distribution $R_\text{d}(\rho)$ as a function of source-detector separation $\rho$. 

\begin{figure}[htb!]
\centering\includegraphics[width=0.45\textwidth]{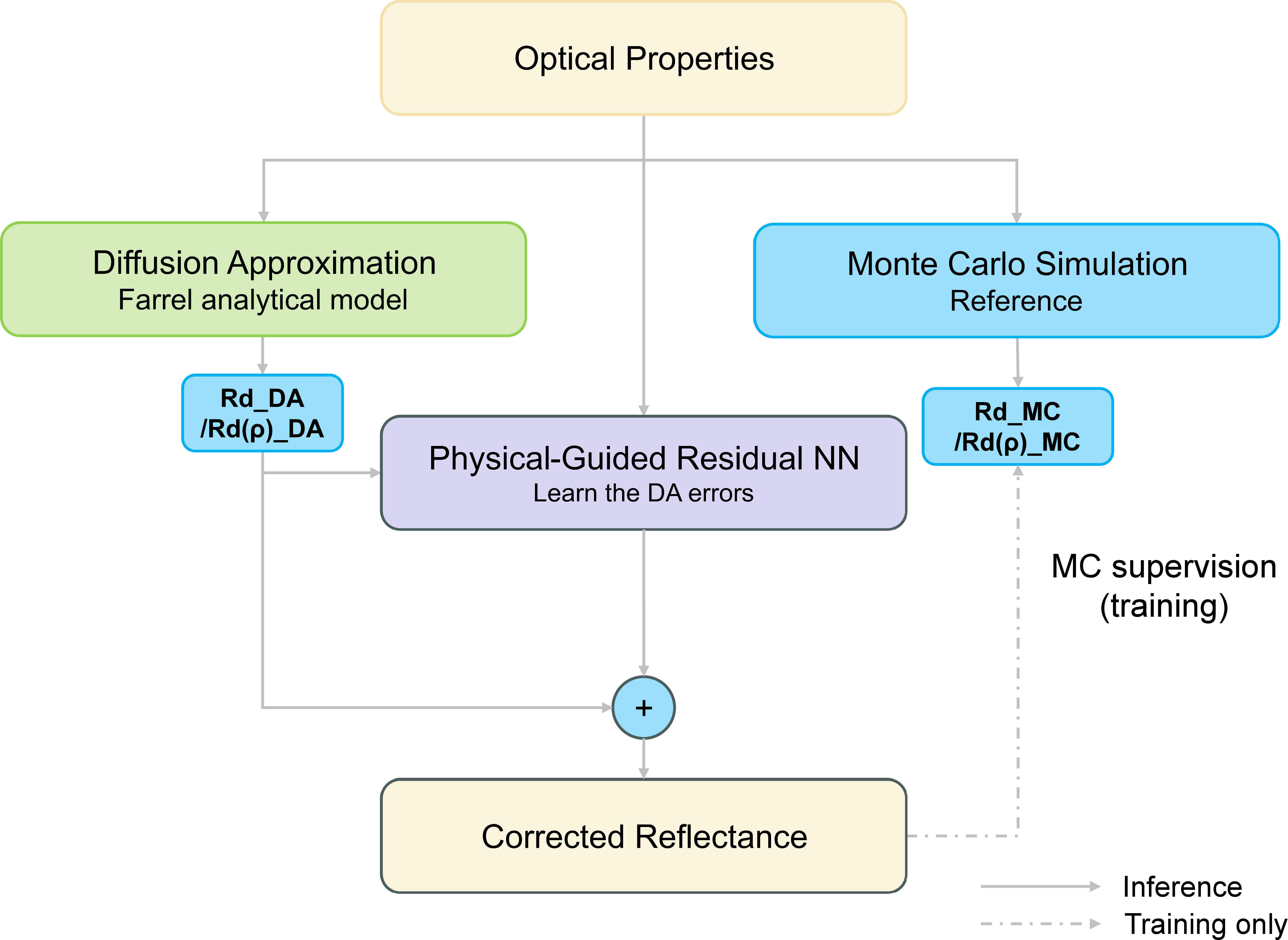}
\caption{The Conceptual Pipeline of the Physical-Guided Correcting Diffusion Approximation.}
\label{fig:pipeline}
\end{figure}

Rather than requiring the neural network to learn the complete mapping from optical properties to reflectance, the proposed framework incorporates the DA prediction as a physics-based prior and learns only the discrepancy between the analytical approximation and the MC reference. The DA-MC discrepancy in spatially resolved reflectance is illustrated conceptually in Figure~\ref{fig:concept} (b), while the complete PGRL workflow is presented in Figure~\ref{fig:pipeline}, highlighting how the DA solution is incorporated both as a physics-based input and as the baseline for the final corrected prediction.

Our proposed model is initialized around an established analytical description of photon transport and is tasked only with learning its systematic correction. Therefore, the formulation is designed to preserve the computational advantages and physical structure of the analytical model while compensating for its regime-dependent deviations from MC. It also reduces the learning burden relative to a fully data-driven surrogate, which is particularly advantageous when the number of available training samples is limited. 

The proposed model was evaluated using MC-generated diffuse reflectance data spanning a broad range of optical properties and compared against the uncorrected DA and direct multilayer perceptron (MLP) baselines. The results demonstrate that PGRL consistently reduces prediction errors across the evaluated optical conditions and training-set sizes.

\section{Materials and Methods}
\subsection{Physics-Guided Residual Learning  Architecture}

Let $\mathbf{x}=(\mu_\text{a},\mu_\text{s},g)$ denote the optical-property vector. 
For each optical property combination, baseline MC simulation provides the reference diffuse reflectance $R_\text{d}^{\text{MC}}$ or $R_\text{d}^{\text{MC}}(\rho)$, whereas the Farrell diffusion model provides a computationally efficient approximation, $
R_\text{d}^\text{DA}=\mathcal{D}(\mathbf{x})$, where $\mathcal{D}(\cdot)$ represents the Farrell diffusion model described below. A conventional data-driven surrogate model directly approximates the complete mapping from the optical properties to the MC reference, $\hat{R}_\text{d}=f_{\theta}(\mathbf{x})$, where $f_{\theta}$ denotes a neural network parameterized by $\theta$. Although this strategy can approximate the input (optical properties) - output (reflectance) relationship when sufficient training data are available, it does not explicitly exploit the physical prior provided by DA. 

\begin{figure}[t!]
\centering\includegraphics[width=0.45\textwidth]{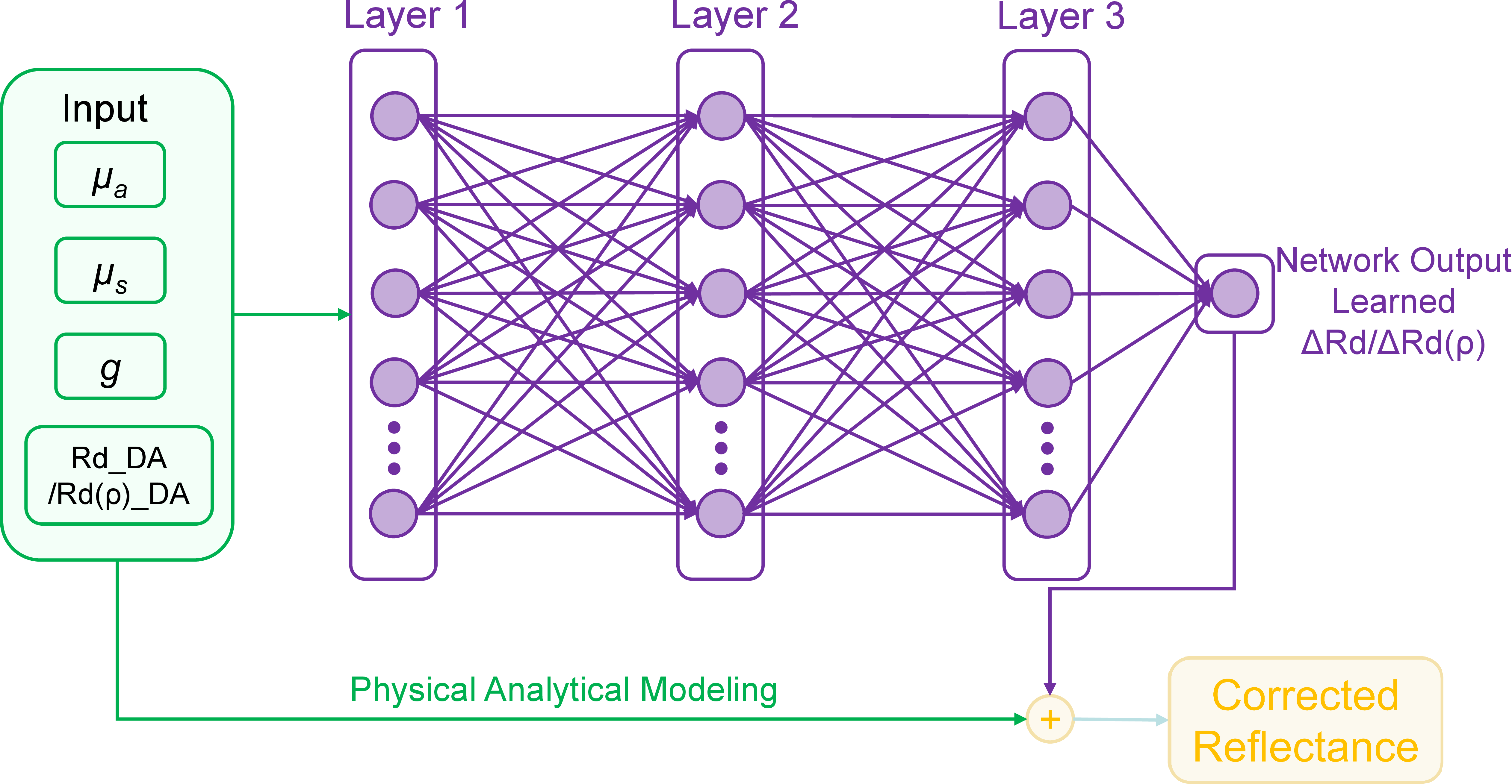}
\caption{The Proposed Physics-Guided Residual Learning architecture}
\label{fig:archi}
\end{figure}

The proposed PGRL architecture is illustrated in Fig.~\ref{fig:archi}. The central idea is to retain the diffusion approximation as an explicit physical baseline and to train the neural network only to recover the discrepancy between the analytical and MC solutions. Accordingly, the corrected reflectance is written as
\begin{equation}
R_\text{d}^\text{MC}
=
R_\text{d}^\text{DA}
+
\Delta \hat{R}_\text{d},
\end{equation}
where $\Delta \hat{R}_\text{d}$ denotes the residual predicted by the network.

This formulation incorporates the DA in two complementary ways. First, $R_\text{d}^\text{DA}$ is supplied to the network together with the optical properties as a physics-guided input, providing an explicit estimate of the dominant reflectance behavior associated with a given combination of $\mu_\text{a}$, $\mu_\text{s}$, and $g$. The network therefore does not need to infer the full reflectance response from the optical properties alone, but instead learns how the analytical prediction should be modified across the investigated optical regimes.

Second, the DA result is retained as the additive baseline for the final output, while the neural network is responsible only for estimating the correction required to approach the MC reference. This residual formulation preserves the physical structure already captured by the analytical model and focuses the learning process on the regime-dependent deviations of the DA. When the predicted correction approaches zero, the model naturally reduces to the original diffusion approximation, whereas nonzero residuals compensate for systematic DA errors where the underlying diffusion assumptions become less accurate.

\subsection{Baseline Dataset and Diffusion Approximation}

The baseline dataset used in this study was generated using an in-house, power-efficient MC model~\cite{Clennell2023NeutralMCEO}. The dataset consist of 828,100 samples, with each sample generated through launching $10^6$ photons to reduce the stochastic noise. The full MC dataset was generated by systematically varying three optical properties of a medium: the absorption coefficient $\mu_\text{a}$, scattering coefficient $\mu_\text{s}$, scattering anisotropy $g$. The absorption coefficient $\mu_\text{a}$, which describes the probability of photon absorption per unit path length, was varied from 1 to 10$\text{\,cm}^{-1}$ with an increment of 0.1$\text{\,cm}^{-1}$. The scattering coefficient $\mu_\text{s}$, which represents the probability of photon scattering per unit path length, ranged from 10 to 1000$\text{\,cm}^{-1}$ in steps of 10$\text{\,cm}^{-1}$. The scattering anisotropy $g$, which quantifies the extent to which photon preserve their original propagation direction after a single scattering event, and is modeled using the Henyey–Greenstein phase function, ranged from 0 to 0.9 with a step size of 0.01. For each combination of three optical properties of a medium, the MC simulation provides both the total diffuse reflectance $R_\text{d}$ and the spatially resolved distribution $R_\text{d}(\rho)$, which serve as the ground truth in the training. The spatially resolved reflectance $R_\text{d}(\rho)$ is represented by a 101-dimensional vector sampled at predefined source-detector separations ranging from 0.0013\,cm to 0.201\,cm.

In this study, we adopt Farrell's steady-state diffusion approximation for a homogeneous semi-infinite medium~\cite{farrell1992diffusion}. In this model, the incident pencil beam is described as a dipole source: an isotropic point source located one transport mean free path below the surface at the depth $z_0 = 1/\mu_\text{t}'$, together with the negative image source at $z_0+2z_b$ to satisfy the extrapolated boundary, where $z_b=2AD$. The spatially resolved reflectance $R_\text{d}^{\text{DA}}(\rho)$ here is given by 
\begin{equation}
\footnotesize
R_\text{d}(\rho) = \frac{a'}{4\pi}
\left[
z_0\!\left(\mu_\text{eff}+\frac{1}{r_1}\right)\frac{e^{-\mu_\text{eff} r_1}}{r_1^{2}}
+ (z_0+2z_b)\!\left(\mu_\text{eff}+\frac{1}{r_2}\right)\frac{e^{-\mu_\text{eff} r_2}}{r_2^{2}}
\right]
\label{eq:rd_rho}
\end{equation}
while the total diffuse reflectance $R_\text{d}^{\text{DA}}$, corresponding to the radial profile integrated over the surface, is given by
\begin{equation}
R_\text{d} = \frac{a'}{2}\,e^{-\mu_\text{eff} z_0}
\left[\,1 + e^{-\frac{4}{3}A\sqrt{3(1-a')}}\,\right]
\label{eq:rd}
\end{equation}


Here, $a' = \mu_\text{s}'/\mu_\text{t}'$ is the transport albedo. $\mu_\text{eff} = \sqrt{\mu_\text{a}/D}$ is the effective attenuation coefficient. $r_1 = \sqrt{z_0^{2}+\rho^{2}}$ and $r_2 = \sqrt{(z_0+2z_b)^{2}+\rho^{2}}$ represent the distances from detector to the isotropic point source and image source. $\mu_\text{s}' = \mu_\text{s}(1-g)$ is the transport scattering coefficient. $D = 1/(3\mu_t')$ is the diffusion coefficient. $\mu_\text{t}' = \mu_\text{a} + \mu_\text{s}'$ is the transport coefficient. The internal reflection factor $A = (1+r_\text{i})/(1-r_\text{i})$ accounts for the refractive-index mismatch at the internal-external interface. $r_\text{i} = 0.668 + 0.0636\,n + 0.710/n - 1.440/n^{2}$ is polynomial in the refractive index $n$. we assume $n=1.5$ for the biological tissues used here. Equations~\ref{eq:rd} and \ref{eq:rd_rho} are evaluated for all the optical property combinations in the MC dataset, producing the physical analytical estimates $R_\text{d}^{\text{DA}}$ and $R_\text{d}^{\text{DA}}(\rho)$ for model training.

\subsection{Training, Evaluation and Ablation study}
To examine the contributions of the two physics-injection mechanisms and the behavior of the model under limited data, a set of controlled comparisons was carried out. The proposed PGRL model was trained on the full MC training dataset to evaluate its overall predictive accuracy. An ablation study was performed using only 1\% of the available training samples to isolate the effects of incorporating the DA prediction as an input feature and retaining it as an additive residual baseline. The four resulting configurations corresponded to the presence or absence of the physics-based input and the presence or absence of the additive physics residual connection. Model performance was also compared with the uncorrected Farrell DA and a direct MLP predictor over multiple training-set sizes.

All trainable models shared a common architecture comprising three fully connected hidden layers, each with 32 neurons and Gaussian error linear unit (GELU) activations~\cite{hendrycks2016gelu}. The MC dataset was partitioned into training, validation, and test subsets in an 8:1:1 ratio. The output dimension was one for total diffuse reflectance $R_\text{d}$ and 101 for spatially resolved reflectance $R_\text{d}(\rho)$, corresponding to the predefined source-detector separations. The final layer of each residual network was initialized to zero so that its initial prediction reproduced the Farrell DA solution. Networks were trained using the AdamW optimizer~\cite{loshchilov2019decoupled}, with a learning rate of $10^{-3}$, weight decay of $10^{-6}$, a batch size of 1024, and a maximum of 500 epochs. Early stopping monitored the validation loss with a patience of 50 epochs, and gradient-norm clipping was applied with a maximum norm of 5~\cite{pascanu2013difficulty}. Each experiment was repeated with five random initialization seeds, and performance is reported as the mean error $\pm$ standard deviation across these runs.

\section{Results and Discussion}

\paragraph{Scalar Diffuse Reflectance}

Table~\ref{tab:test_performance1} summarizes the test-set performance for
scalar diffuse reflectance using the full training dataset. The Farrell DA
produces an RMSE of $1.77\times10^{-2}$, whereas the direct MLP reduces the
error to $(5.20\pm0.28)\times10^{-4}$. The proposed PGRL model further
reduces the RMSE to $(2.85\pm0.05)\times10^{-4}$, corresponding to
improvements by factors of approximately 62.1 and 1.82 relative to the
Farrell DA and direct MLP, respectively.

\begin{table}[h]
\centering
\caption{Prediction performance for scalar diffuse reflectance using the
full training set.}
\label{tab:test_performance1}
\small
\begin{tabular}{lcc}
\hline
Model & RMSE & $R^2$ \\
\hline
Farrell DA &
$1.77\times10^{-2}$ &
0.988689 \\

MLP &
$(5.20\pm0.28)\times10^{-4}$ &
0.999990 \\

\textbf{PGRL} &
$(2.85\pm0.05)\times10^{-4}$ &
0.999997 \\
\hline
\end{tabular}
\end{table}

\begin{figure*}[!t]
\centering

\begin{minipage}[t]{0.47\textwidth}
    \centering
    \includegraphics[width=0.92\textwidth]{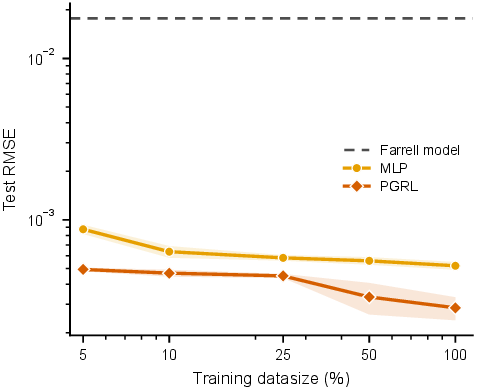}
    \vspace{1mm}
    \makebox[\textwidth][c]{\small (a)}
\end{minipage}
\hfill
\begin{minipage}[t]{0.47\textwidth}
    \centering
    \includegraphics[width=0.92\textwidth]{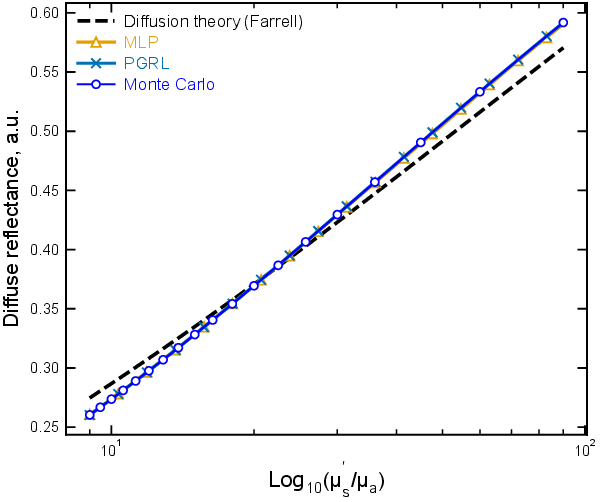}
    \vspace{1mm}
    \makebox[\textwidth][c]{\small (b)}
\end{minipage}

\vspace{3mm}

\begin{minipage}[t]{0.47\textwidth}
    \centering
    \includegraphics[width=0.92\textwidth]{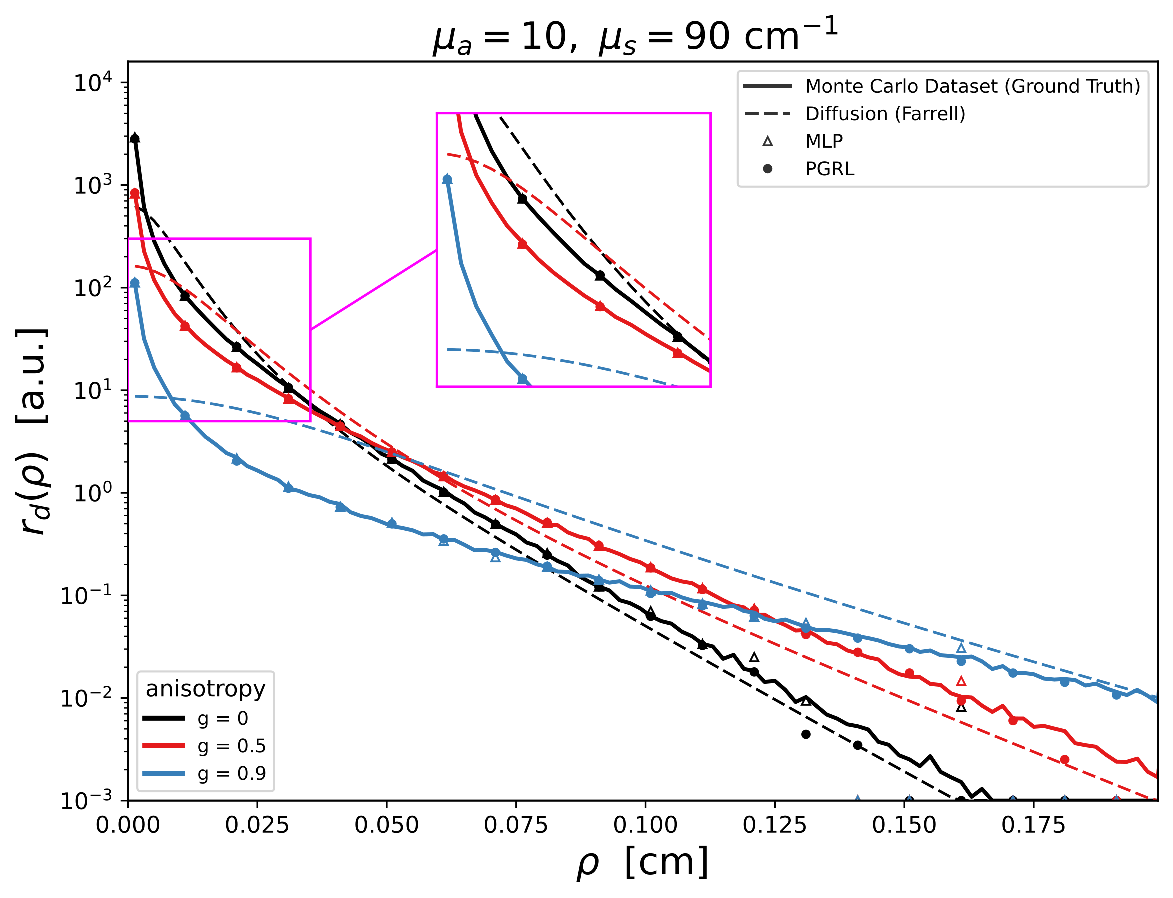}
    \vspace{1mm}
    \makebox[\textwidth][c]{\small (c)}
\end{minipage}
\hfill
\begin{minipage}[t]{0.47\textwidth}
    \centering
    \includegraphics[width=0.92\textwidth]{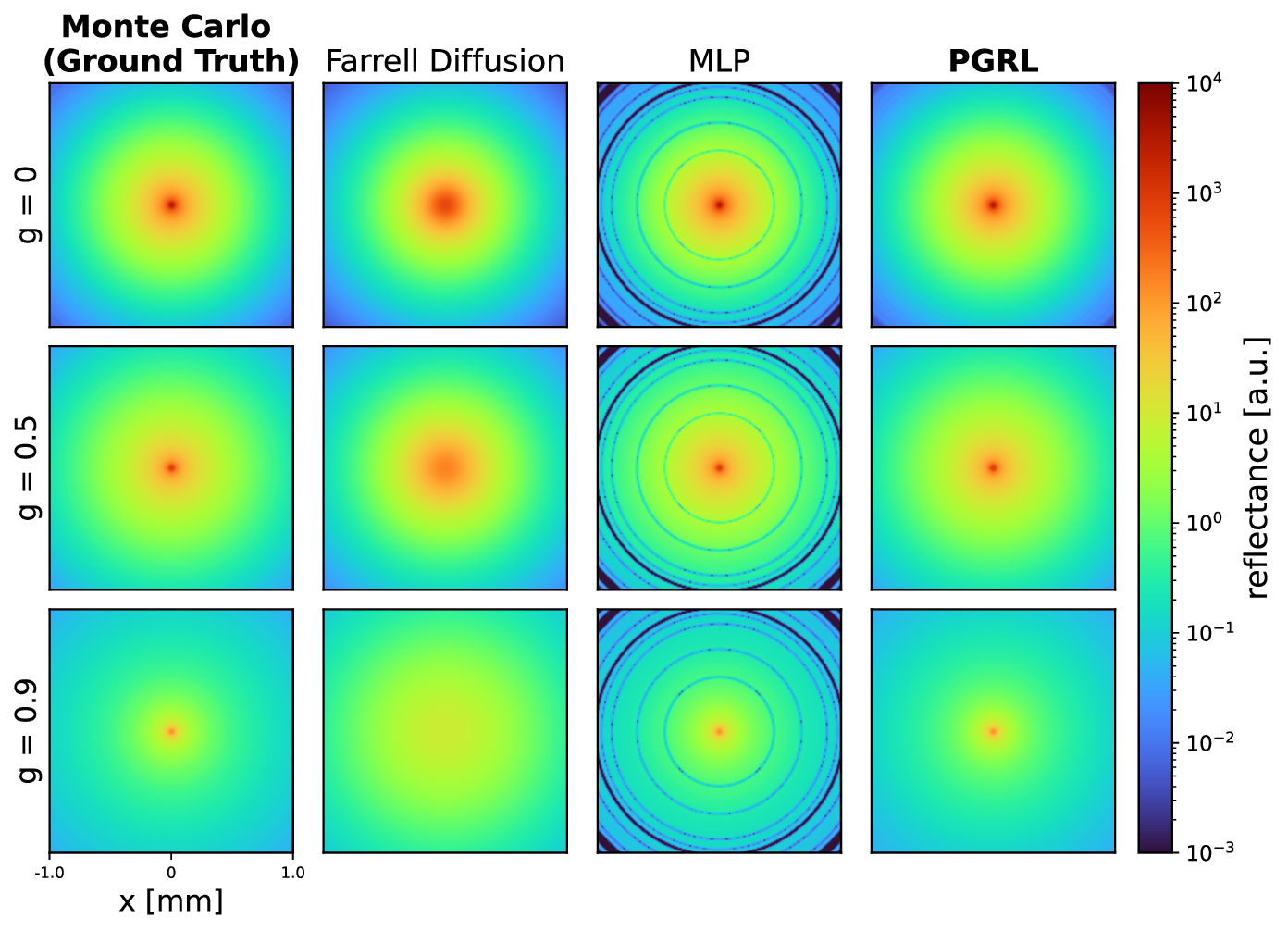}
    \vspace{1mm}
    \makebox[\textwidth][c]{\small (d)}
\end{minipage}

\caption{Performance of the proposed physics-guided residual learning
(PGRL) framework for scalar and spatially resolved diffuse reflectance.
(a) Test-set RMSE of the direct MLP and PGRL models as the amount of
available MC training data is varied.
(b) Comparison between predicted scalar diffuse reflectance
$R_\mathrm{d}$ and the MC reference across the investigated
optical-property space.
(c) Spatially resolved reflectance profiles $R_\mathrm{d}(\rho)$ for
$\mu_\mathrm{a}=10~\mathrm{cm}^{-1}$,
$\mu_\mathrm{s}=90~\mathrm{cm}^{-1}$, and
$g=0$, $0.5$, and $0.9$.
(d) Corresponding two-dimensional radially symmetric reflectance maps.
The PGRL model corrects systematic deviations of the Farrell diffusion
approximation while retaining the smooth spatial structure of the
MC reference.}
\label{fig:results}
\end{figure*}

Figure~\ref{fig:results}(b) shows the relationship between the predicted
and MC-reference values across the investigated optical-property space.
The DA error is clearly regime dependent: the analytical model tends to
overestimate reflectance toward stronger absorption and underestimate it
toward stronger scattering. In contrast, the PGRL predictions remain close
to the MC reference throughout these regions. This behavior indicates that
the learned residual primarily compensates for systematic, regime-dependent
deviations of the diffusion approximation while retaining the overall
physical trend provided by the analytical model.

The advantage of incorporating the analytical solution becomes more
pronounced when the amount of MC supervision is reduced.
Table~\ref{tab:ablation_study1} presents the ablation study using only
1\% of the available training data. Under this condition, the direct MLP
reaches an RMSE of $(4.17\pm0.41)\times10^{-3}$. Supplying the Farrell
solution only as an additional network input reduces the error by
approximately 73\% to $(1.13\pm0.07)\times10^{-3}$, while using the DA
only as an additive residual baseline produces a comparable RMSE of
$(1.17\pm0.28)\times10^{-3}$. Combining both mechanisms in the complete
PGRL formulation further decreases the RMSE to
$(5.54\pm1.50)\times10^{-4}$, corresponding to an approximately 87\%
reduction relative to the direct MLP.

\begin{table}[ht]
\centering
\caption{Ablation study of the proposed physics-guided components for
scalar diffuse reflectance using 1\% of the training data.}
\label{tab:ablation_study1}
\small
\begin{tabular}{lcc}
\hline
Model & RMSE & $R^2$ \\
\hline
Farrell DA &
$1.77\times10^{-2}$ &
0.988689 \\

MLP &
$(4.17\pm0.41)\times10^{-3}$ &
$0.999368\pm0.000126$ \\

MLP + Farrell Input &
$(1.13\pm0.07)\times10^{-3}$ &
$0.999954\pm0.000006$ \\

Residual MLP &
$(1.17\pm0.28)\times10^{-3}$ &
$0.999948\pm0.000026$ \\

\textbf{PGRL} &
$(0.554\pm0.15)\times10^{-3}$ &
$0.999989\pm0.000001$ \\
\hline
\end{tabular}
\end{table}

These results demonstrate that the analytical DA estimate contributes useful
information through both pathways: as a physics-guided input and as the
baseline of the residual prediction. The same trend is observed in the
training-set sensitivity analysis shown in Fig.~\ref{fig:results}(a), where
PGRL consistently maintains a lower RMSE than the direct MLP as the amount
of MC training data is reduced. This persistent advantage supports the
central premise of the proposed framework: learning the discrepancy between
an existing analytical solution and the MC reference is more data efficient
than learning the complete optical-property-to-reflectance mapping directly
from data.

\paragraph{Spatially Resolved Diffuse Reflectance}

The spatially resolved reflectance $R_\mathrm{d}(\rho)$ spans approximately
four orders of magnitude across the 101 source--detector separations.
Prediction accuracy was therefore evaluated using the root mean squared
logarithmic error (RMSLE). As summarized in
Table~\ref{tab:test_performance2}, the Farrell DA produces an RMSLE of
$0.202240$, while the direct MLP achieves $0.233\pm0.230$. In contrast,
PGRL reduces the RMSLE to $0.014\pm0.002$, corresponding to improvements
by factors of approximately 14.4 and 16.6 relative to the Farrell DA and
direct MLP, respectively.

\begin{table}[ht]
\centering
\caption{Prediction performance for spatially resolved diffuse reflectance
using the full training set.}
\label{tab:test_performance2}
\small
\begin{tabular}{lcc}
\hline
Model & RMSLE & $R^2$ \\
\hline
Farrell DA &
$0.202240$ &
$0.964155$ \\

MLP &
$0.233\pm0.230$ &
$0.997102\pm0.006287$ \\

\textbf{PGRL} &
$0.014\pm0.002$ &
$0.999890\pm0.000029$ \\
\hline
\end{tabular}
\end{table}

The large variation in the MLP RMSLE across random initializations further
indicates that direct learning of the complete spatial reflectance profile is
considerably less stable than the residual formulation. In PGRL, the network
does not need to reconstruct the entire 101-dimensional spatial distribution
from the optical properties alone; instead, it begins from the physically
structured DA profile and learns only its discrepancy from the MC reference.

The low-data ablation study in Table~\ref{tab:ablation_study2} reveals a
different contribution pattern from that observed for scalar reflectance.
Using only 1\% of the training data, the direct MLP produces an RMSLE of
$0.454\pm0.313$. Supplying the Farrell prediction only as an additional
input provides a modest improvement to $0.427\pm0.277$. In contrast,
retaining the Farrell solution as the additive residual baseline reduces the
RMSLE substantially to $0.049\pm0.003$. The complete PGRL formulation,
which combines both the physics-guided input and residual baseline, further
reduces the RMSLE to $0.033\pm0.003$.

\begin{table}[htb]
\centering
\caption{Ablation study of the proposed physics-guided components for
spatially resolved diffuse reflectance using 1\% of the training data.}
\label{tab:ablation_study2}
\small
\begin{tabular}{lcc}
\hline
Model & RMSLE & $R^2$ \\
\hline
Farrell DA &
$0.202240$ &
$0.964155$ \\

MLP &
$0.454\pm0.313$ &
$0.593854\pm0.051309$ \\

MLP + Farrell Input &
$0.427\pm0.277$ &
$0.658110\pm0.102953$ \\

Residual MLP &
$0.049\pm0.003$ &
$0.975164\pm0.000746$ \\

\textbf{PGRL} &
$0.033\pm0.003$ &
$0.977060\pm0.001008$ \\
\hline
\end{tabular}
\end{table}

These results indicate that the residual connection plays a particularly
important role for spatially resolved prediction. The analytical DA profile
already contains the dominant radial structure of the reflectance
distribution, allowing the network to focus on learning the systematic
difference between the analytical approximation and the MC reference.
Providing the DA prediction additionally as an input offers a further
improvement by allowing the network to relate the magnitude and structure
of this discrepancy to the underlying optical properties.

Representative spatial predictions are shown in
Fig.~\ref{fig:results}(c,d) for
$\mu_\mathrm{a}=10~\mathrm{cm}^{-1}$,
$\mu_\mathrm{s}=90~\mathrm{cm}^{-1}$, and anisotropy factors
$g=0$, $0.5$, and $0.9$.
Figure~\ref{fig:results}(c) compares the radial profiles
$R_\mathrm{d}(\rho)$. Across all three anisotropy conditions, PGRL closely
reproduces the MC radial decay, whereas the Farrell DA exhibits systematic
deviations and the direct MLP shows less stable spatial behavior.

The corresponding two-dimensional representations are shown in
Fig.~\ref{fig:results}(d). Because the underlying problem is radially
symmetric, deviations in the predicted radial profiles appear as concentric
structures in the reconstructed maps. Pronounced ring-like artifacts are
visible in the direct MLP predictions, whereas the PGRL maps retain the
smooth spatial structure of the MC reference. This suggests that using the
analytical DA solution as a residual baseline not only reduces the numerical
prediction error but also helps preserve the physically expected spatial
structure of diffuse reflectance.

\section{Conclusion}
In this study, we introduced a physics-guided residual learning framework that improves analytical photon transport predictions by learning systematic corrections to the diffusion approximation from MC simulations. Incorporating the analytical solution as both a network input and an explicit residual baseline improves predictions of total and spatially resolved diffuse reflectance, with benefits maintained even when using only 1\% of the available MC training data. Ablation studies highlight the importance of the analytical baseline for accurate spatial predictions. Together, these results demonstrate that established physical models can reduce the learning burden and improve both prediction accuracy and data efficiency/throughput.

More broadly, presented approach offers a way to extend the practical utility of analytical models into regimes where their underlying approximations become less accurate. By retaining the analytical description and learning its discrepancies from a higher-fidelity reference, physics-guided residual learning connects tractable physical theory with data-driven correction. The same principle could be applied beyond diffusion theory and photon transport wherever an analytical or semi-analytical model captures the dominant behavior but requires systematic refinement.

\section{Disclosures}
The authors declare no conflicts of interest.

\bibliography{ref}

\bibliographyfullrefs{ref}

\end{document}